\documentclass[usegraphicx]{mn2e}

\title[A simple and accurate approximation for $Q$]
      {A simple and accurate approximation for the $Q$ stability parameter in
       multi-component and realistically thick discs}

\author[A. B. Romeo and N. Falstad]
       {Alessandro B. Romeo\thanks{E-mail: romeo@chalmers.se}
        and Niklas Falstad\\
        Department of Earth and Space Sciences,
        Chalmers University of Technology,
        Onsala Space Observatory,
        SE-43992 Onsala, Sweden}

\begin{document}

\date{Accepted 2013 May 8.
      Received 2013 April 22; in original form 2013 February 18}

\pagerange{\pageref{firstpage}--\pageref{lastpage}}

\pubyear{2013}

\maketitle

\label{firstpage}

\begin{abstract}
In this paper, we propose a $Q$ stability parameter that is more realistic
than those commonly used, and is easy to evaluate [see Eq.\ (19)].  Using our
$\mathcal{Q}_{N}$ parameter, you can take into account several stellar and/or
gaseous components as well as the stabilizing effect of disc thickness, you
can predict which component dominates the local stability level, and you can
do all that simply and accurately.  To illustrate the strength of
$\mathcal{Q}_{N}$, we analyse the stability of a large sample of spirals from
The H\,\textsc{i} Nearby Galaxy Survey (THINGS), treating stars,
H\,\textsc{i} and $\mathrm{H}_{2}$ as three distinct components.  Our
analysis shows that $\mathrm{H}_{2}$ plays a significant role in disc
(in)stability even at distances as large as half the optical radius.  This is
an important aspect of the problem, which was missed by previous
(two-component) analyses of THINGS spirals.  We also show that H\,\textsc{i}
plays a negligible role up to the edge of the optical disc; and that the
stability level of THINGS spirals is, on average, remarkably flat and well
above unity.
\end{abstract}


\section{INTRODUCTION}

Today, several decades after the pioneering work of Lin \& Shu (1966) and the
seminal papers by Jog \& Solomon (1984a,\,b), it is widely accepted that
stars and cold interstellar gas have an important interplay in the
gravitational instability of galactic discs.  The gravitational coupling
between stars and gas does not alter the form of the local axisymmetric
stability criterion, $Q\geq1$ (Toomre 1964), but makes the $Q$ stability
parameter dependent on the radial velocity dispersions and surface mass
densities of the two components (Bertin \& Romeo 1988; Elmegreen 1995; Jog
1996; Rafikov 2001; Shen \& Lou 2003).  The value of $Q$ is also affected by
other factors, such as the vertical structure of the disc (Shu 1968; Romeo
1990, 1992, 1994; Elmegreen 2011; Romeo \& Wiegert 2011), gas turbulence
(Hoffmann \& Romeo 2012; Shadmehri \& Khajenabi 2012), and gas dissipation
(Elmegreen 2011).  Comprehensive analyses have shown that the two-component
$Q$ parameter has a large impact on the dynamics and evolution of spiral
structure in galaxies (see Bertin \& Lin 1996), and is also a useful
diagnostic for exploring the link between disc instability and star formation
(Leroy et al.\ 2008).

Romeo \& Wiegert (2011) introduced a simple and accurate approximation for
the two-component $Q$ parameter, which takes into account the stabilizing
effect of disc thickness and predicts whether the local stability level is
dominated by stars or gas.  The Romeo-Wiegert approximation has been used for
investigating the evolution of gravitationally unstable discs (Cacciato et
al.\ 2012; Forbes et al.\ 2012), the spiral structure of NGC 5247 (Khoperskov
et al.\ 2012), the dynamical link between dark matter and H\,\textsc{i} in
nearby galaxies (Meurer et al.\ 2013), as well as the link between disc
stability and the relative distributions of stars, gas and star formation
(Zheng et al.\ 2013).  Forbes et al.\ (2012) concluded that the Romeo-Wiegert
approximation is much faster to use than the $Q$ stability parameter of
Rafikov (2001): it speeds up their disc-evolution code by as much as one or
two orders of magnitude!  This is simply because such an approximation
estimates $Q$ analytically, without the need to minimize the dispersion
relation over all wavenumbers, as is usually done.

A fundamental problem that must be faced when analysing the stability of
galactic discs is how to represent their complex structure using only two
components.  The results of such analyses are indeed very sensitive to the
choice of the gaseous 1D velocity dispersion, $\sigma_{\mathrm{g}}$: the
colder the gas, the stronger its impact on the stability of the disc (e.g.,
Jog \& Solomon 1984a; Bertin \& Romeo 1988).  Choosing
$\sigma_{\mathrm{g}}\approx6\;\mbox{km\,s}^{-1}$ will represent molecular gas
well (Wilson et al.\ 2011), but will overestimate the contribution of atomic
gas and make the disc more unstable than it actually is.  Vice versa,
choosing $\sigma_{\mathrm{g}}\approx11\;\mbox{km\,s}^{-1}$ will represent
H\,\textsc{i} well (Leroy et al.\ 2008), but will underestimate the
contribution of $\mathrm{H}_{2}$ and stabilize the disc artificially.
Intermediate values or more elaborate choices of $\sigma_{\mathrm{g}}$ will
still not solve the problem.  This motivates the use of a proper
multi-component $Q$ parameter.

One of the first papers that discussed the gravitational instability of
multi-component discs dates back to Morozov (1981).  This author derived a
dispersion relation that is valid for infinitesimally thin discs made of gas
and $N_{\star}$ stellar components.  He also calculated the stability
criterion for $N_{\star}=2$.  Rafikov (2001) derived a stability criterion
that is valid for any $N_{\star}$ and can easily be expressed in the usual
form $Q\geq1$.%
\footnote{Another (unpublished) stability analysis of $N$-component discs was
  made by Romeo (1985), pp.\ 140--145 and 215--216.}

In this paper, we introduce a new $Q$ stability parameter, which has the same
strong advantages as the Romeo-Wiegert approximation, and which is applicable
to fully multi-component and realistically thick discs [see Eq.\ (19)].  We
also show how to use our $\mathcal{Q}_{N}$ parameter for analysing the
stability of galactic discs, and illustrate the strength of a multi-component
analysis.  We do so using a large sample of spirals from The H\,\textsc{i}
Nearby Galaxy Survey (THINGS), and treating stars, H\,\textsc{i} and
$\mathrm{H}_{2}$ as three distinct components.

The rest of the paper is organized as follows.  In Sect.\ 2, we review the
basic case of two-component discs and further motivate the need for a
multi-component analysis (Sect.\ 2.1), we present our $\mathcal{Q}_{N}$
parameter (Sects 2.2--2.4), and we analyse the stability of THINGS spirals
(Sect.\ 2.5).  In Sect.\ 3, we discuss the weaknesses of $\mathcal{Q}_{N}$,
which are common to all $Q$ parameters and stability criteria quoted here.
In Sect.\ 4, we draw the conclusions.

\section{STAR-GAS INSTABILITIES AND THE $Q$ DIAGNOSTIC}

\subsection{Two components \ldots\ or more?  More!}

Let us first discuss the case of two-component and infinitesimally thin
discs, which is fundamental to a proper understanding of Sects 2.2--2.5.  It
is well known that the stability properties of such discs are determined by
five basic quantities: the epicyclic frequency, $\kappa$, the stellar and
gaseous surface densities, $\Sigma_{\star}$ and $\Sigma_{\mathrm{g}}$, the
stellar radial velocity dispersion, $\sigma_{R\star}$, and the gaseous 1D
velocity dispersion, $\sigma_{\mathrm{1D\,g}}$ (e.g., Lin \& Shu 1966;
Rafikov 2001).  The last two quantities reflect an important dynamical
difference between stars and cold interstellar gas.  The stellar component is
collisionless so its velocity dispersion is anisotropic, while gas is
collisional and has an isotropic velocity dispersion (see, e.g., Binney \&
Tremaine 2008).  Hereafter we will simplify the notation and denote the
relevant velocity dispersions of the two components with $\sigma_{\star}$ and
$\sigma_{\mathrm{g}}$.

Lin \& Shu (1966) and Rafikov (2001) took those facts into account by
treating stars as a kinetic component and gas as a fluid (see also Shu 1968).
The resulting local axisymmetric stability criteria are equivalent because
they are based on the same dispersion relation, $\omega^{2}(k)$, and because
they are derived by imposing that $\omega^{2}(k)\geq0$ for all $k$.  However,
Rafikov's stability criterion is simpler and more used (e.g., Dalcanton et
al.\ 2004; Li et al.\ 2005, 2006; Kim \& Ostriker 2007; Yang et al.\ 2007;
Yim et al.\ 2011).  Such a criterion can be written as
$Q_{\mathrm{Rk}}\geq1$, where
\begin{equation}
\frac{1}{Q_{\mathrm{Rk}}}=
\max\{\mathrm{SC}_{\mathrm{Rk}}(K)\}\,,
\end{equation}
\begin{equation}
\mathrm{SC}_{\mathrm{Rk}}(K)=
\frac{1}{Q_{\star}}\,
\frac{2\left[1-\mathrm{e}^{-K^{2}}I_{0}\left(K^{2}\right)\right]}{K}+
\frac{1}{Q_{\mathrm{g}}}\,
\frac{2Ks}{1+K^{2}s^{2}}\,,
\end{equation}
\begin{equation}
K=
k\,\frac{\sigma_{\star}}{\kappa}\,,
\end{equation}
\begin{equation}
Q_{\star}=
\frac{\kappa\sigma_{\star}}{\pi G\Sigma_{\star}}\,,\;\;\;\;\;
Q_{\mathrm{g}}=
\frac{\kappa\sigma_{\mathrm{g}}}{\pi G\Sigma_{\mathrm{g}}}\,,
\end{equation}
\begin{equation}
s=
\frac{\sigma_{\mathrm{g}}}{\sigma_{\star}}\,.
\end{equation}
In these equations, $Q_{\mathrm{Rk}}$ is \underline{R}afikov's
\underline{k}inetic-fluid $Q$ parameter, $\mathrm{SC}_{\mathrm{Rk}}(K)$ is
the related stability curve, $K$ is the radial wavenumber of the perturbation
expressed in dimensionless form, $Q_{\star}$ and $Q_{\mathrm{g}}$ are the
stellar and gaseous Toomre parameters, and $I_{0}$ denotes the modified
Bessel function of the first kind and order zero.

%%%%%%%%%%%%%%%%%%%%%%%%%%%%%%%%%%%%%%%%%%%%%%%%%%%%%%%%%%%%%%%%%%%%%%%%%%%%%
\begin{figure}
\includegraphics[scale=1.]{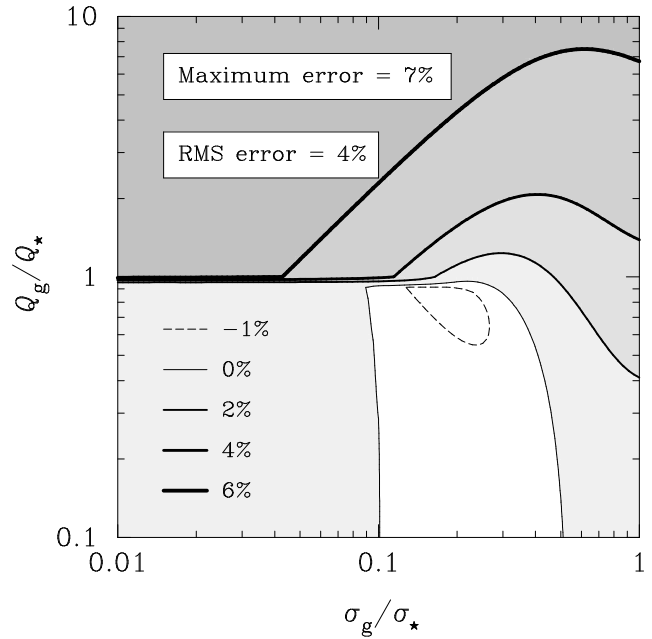}
\caption{Accuracy of the two-fluid stability criterion.  The curves shown are
  the contour lines of the relative error
  $(Q_{\mathrm{Rf}}-Q_{\mathrm{Rk}})/Q_{\mathrm{Rk}}$, where
  $Q_{\mathrm{Rf}}$ and $Q_{\mathrm{Rk}}$ are the fluid-fluid and
  kinetic-fluid $Q$ parameters of Rafikov (2001).  In addition, $Q_{\star}$
  and $Q_{\mathrm{g}}$ are the stellar and gaseous Toomre parameters,
  $\sigma_{\star}$ and $\sigma_{\mathrm{g}}$ are the relevant velocity
  dispersions of the two components.}
\end{figure}
%%%%%%%%%%%%%%%%%%%%%%%%%%%%%%%%%%%%%%%%%%%%%%%%%%%%%%%%%%%%%%%%%%%%%%%%%%%%%

Bertin \& Romeo (1988), Elmegreen (1995), Jog (1996), and again Rafikov
(2001) adopted a less rigorous, but more straightforward approach: they
treated the stellar component as a fluid, with sound speed equal to
$\sigma_{\star}$.  The resulting two-fluid stability criteria are equivalent,
apart from the different parametrizations used, because they are based on the
same dispersion relation (Jog \& Solomon 1984a).  However, even in this case,
Rafikov's criterion is simpler and is becoming more and more widely used
(e.g., Leroy et al.\ 2008; Robertson \& Kravtsov 2008; Dekel et al.\ 2009;
Mastropietro et al.\ 2009; Ceverino et al.\ 2010; Westfall et al.\ 2011;
Elson et al.\ 2012; Watson et al.\ 2012; Williamson \& Thacker 2012).  Such a
criterion can be written as $Q_{\mathrm{Rf}}\geq1$, where
\underline{R}afikov's \underline{f}luid-fluid $Q$ parameter can be computed
by maximizing the related stability curve over all radial wavenumbers:
\begin{equation}
\frac{1}{Q_{\mathrm{Rf}}}=
\max\{\mathrm{SC}_{\mathrm{Rf}}(K)\}\,,
\end{equation}
\begin{equation}
\mathrm{SC}_{\mathrm{Rf}}(K)=
\frac{1}{Q_{\star}}\,
\frac{2K}{1+K^{2}}+
\frac{1}{Q_{\mathrm{g}}}\,
\frac{2Ks}{1+K^{2}s^{2}}\,.
\end{equation}
What is the accuracy of the two-fluid stability criterion?  Bertin \& Romeo
(1988) compared the fluid-fluid and kinetic-fluid marginal stability curves
in a representative set of cases, and showed that the differences are small
(see their fig.\ 2).  Rafikov (2001) carried out a more detailed analysis and
got similar results (see his fig.\ 3).  However, he did not evaluate the
relative error $(Q_{\mathrm{Rf}}-Q_{\mathrm{Rk}})/Q_{\mathrm{Rk}}$, which
depends on $s$ and
\begin{equation}
q=
\frac{Q_{\mathrm{g}}}{Q_{\star}}\,,
\end{equation}
and which is a useful piece of information for assessing the accuracy of
$Q_{\mathrm{Rf}}$.  We do this in Fig.\ 1.  Our contour map shows that such
error is less than 7\% and has a root-mean-square value of 4\%.  This means
that $Q_{\mathrm{Rf}}$ and $Q_{\mathrm{Rk}}$ are practically equivalent,
since the parameters $s$ and $q$ are themselves subject to observational
uncertainties.

%%%%%%%%%%%%%%%%%%%%%%%%%%%%%%%%%%%%%%%%%%%%%%%%%%%%%%%%%%%%%%%%%%%%%%%%%%%%%
\begin{figure}
\includegraphics[scale=1.]{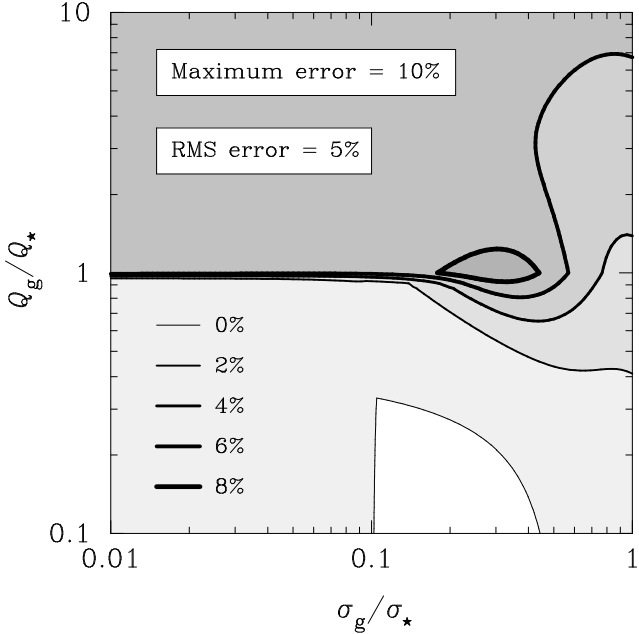}
\caption{Accuracy of the Romeo-Wiegert approximation.  The curves shown are
  the contour lines of the relative error
  $(Q_{\mathrm{RW}}-Q_{\mathrm{Rk}})/Q_{\mathrm{Rk}}$, where
  $Q_{\mathrm{RW}}$ is the two-component $Q$ parameter of Romeo \& Wiegert
  (2011), and $Q_{\mathrm{Rk}}$ is the kinetic-fluid $Q$ parameter of Rafikov
  (2001).  The rest of the notation is the same as in Fig.\ 1.}
\end{figure}
%%%%%%%%%%%%%%%%%%%%%%%%%%%%%%%%%%%%%%%%%%%%%%%%%%%%%%%%%%%%%%%%%%%%%%%%%%%%%

Romeo \& Wiegert (2011) showed that the two-fluid $Q$ parameter can be
accurately estimated without the usual maximization (or minimization)
procedure:
\begin{equation}
\frac{1}{Q_{\mathrm{RW}}}=
\left\{\begin{array}{ll}
       {\displaystyle\frac{W}{Q_{\star}}+\frac{1}{Q_{\mathrm{g}}}}
                       & \mbox{if\ }Q_{\star}\geq Q_{\mathrm{g}}\,, \\
                       &                                            \\
       {\displaystyle\frac{1}{Q_{\star}}+\frac{W}{Q_{\mathrm{g}}}}
                       & \mbox{if\ }Q_{\mathrm{g}}\geq Q_{\star}\,;
       \end{array}
\right.
\end{equation}
\begin{equation}
W=
\frac{2\sigma_{\star}\sigma_{\mathrm{g}}}
     {\sigma_{\star}^{2}+\sigma_{\mathrm{g}}^{2}}\,.
\end{equation}
Although very recent, such an approximation is already frequently used (e.g.,
Cacciato et al.\ 2012; Forbes et al.\ 2012; Hoffmann \& Romeo 2012;
Khoperskov et al.\ 2012; Meurer et al.\ 2013; Zheng et al.\ 2013).  Fig.\ 2
shows that $Q_{\mathrm{RW}}$ is remarkably accurate even with respect to the
kinetic-fluid $Q$ parameter.  In fact, the relative error
$(Q_{\mathrm{RW}}-Q_{\mathrm{Rk}})/Q_{\mathrm{Rk}}$ is below 10\% and has a
root-mean-square value of 5\%.  Thus $Q_{\mathrm{RW}}$ is a faster and
physically equivalent alternative to $Q_{\mathrm{Rf}}$ or $Q_{\mathrm{Rk}}$.

%%%%%%%%%%%%%%%%%%%%%%%%%%%%%%%%%%%%%%%%%%%%%%%%%%%%%%%%%%%%%%%%%%%%%%%%%%%%%
\begin{figure*}
\includegraphics[scale=.96]{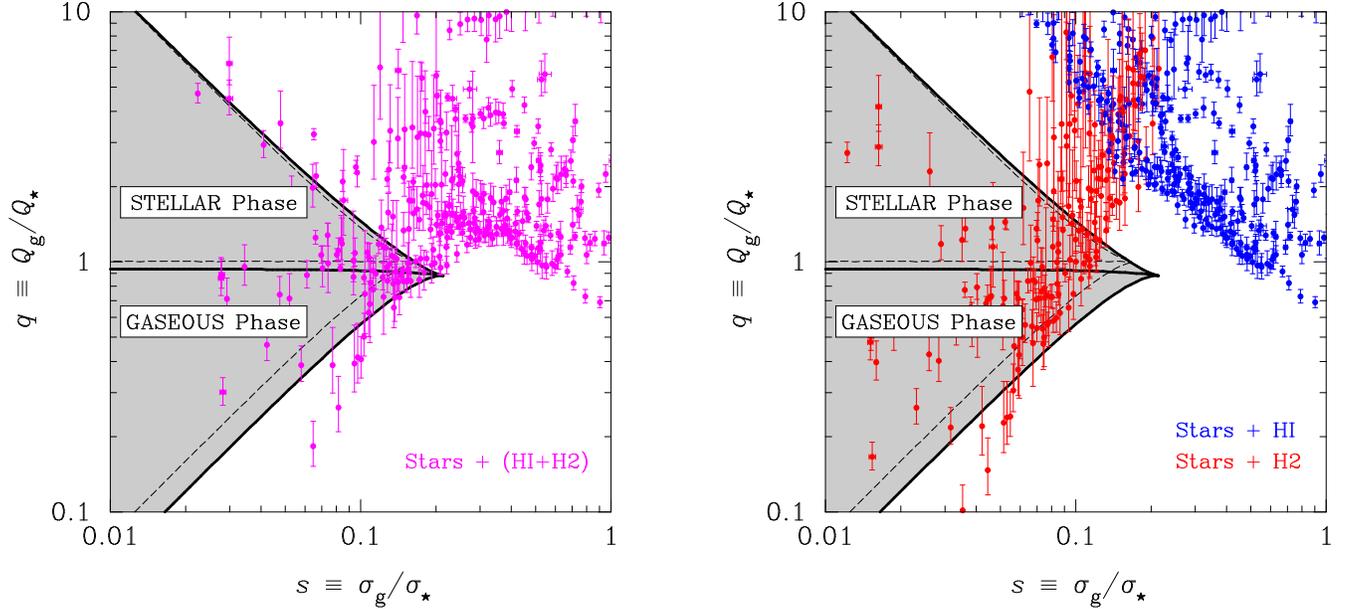}
\caption{Why a multi-component analysis is needed: the parameter plane of
  star-gas instabilities populated by THINGS spirals.  The cases illustrated
  in the two panels are stars plus gas (left), and stars plus H\,\textsc{i}
  or $\mathrm{H}_{2}$ (right).  The galaxy data are from Leroy et
  al.\ (2008), $Q_{\star}$ and $Q_{\mathrm{g}}$ are the stellar and gaseous
  Toomre parameters, $\sigma_{\star}$ and $\sigma_{\mathrm{g}}$ are the
  relevant velocity dispersions of the two components.  The shaded part of
  the $(s,q)$ plane represents the condition for star-gas decoupling, with
  the stellar component treated as collisionless (thick solid lines) or
  collisional (thin dashed lines).  This is the `two-phase region' discussed
  in the text.  The boundaries of this region and the transition line
  intersect at $(s,q)\simeq(0.21,0.88)$ in the collisionless case, and at
  $(s,q)\simeq(0.17,1)$ in the collisional case.}
\end{figure*}
%%%%%%%%%%%%%%%%%%%%%%%%%%%%%%%%%%%%%%%%%%%%%%%%%%%%%%%%%%%%%%%%%%%%%%%%%%%%%

To understand the weaknesses of a two-component analysis, let us see how
spiral galaxies populate the parameter plane of star-gas instabilities.  We
use a sample of twelve nearby star-forming spirals from THINGS, previously
analysed by Leroy et al.\ (2008) and Romeo \& Wiegert (2011): NGC 628, 2841,
3184, 3198, 3351, 3521, 3627, 4736, 5055, 5194, 6946 and 7331.  For each
galaxy of this sample, we compute the radial profiles $s=s(R)$ and $q=q(R)$,
and hence the track left by the galaxy in the $(s,q)$ plane.  The result for
the whole sample is shown in the left panel of Fig.\ 3.  Note that 20\% of
the data fall within the shaded part of the $(s,q)$ plane, which represents
the condition for star-gas decoupling.  In this region,
$\mathrm{SC}_{\mathrm{Rk}}(K)$ has two maxima: one at small $K$, where the
response of the stellar component peaks; and the other at large $K$, where
gas dominates.  In the `stellar phase', the maximum at small $K$ is higher
than the other one, and therefore it controls the onset of disc instability.
Vice versa, in the `gaseous phase', it is the maximum at large $K$ that
determines $Q_{\mathrm{Rk}}$.  The two-fluid counterpart of this region (thin
dashed lines) is also populated by 20\% of the data (Romeo \& Wiegert 2011).
In the rest of the parameter plane, the dynamical responses of the two
components are strongly coupled and peak at a single wavelength.

The analysis of THINGS spirals carried out above treats the interstellar
medium (ISM) as a single component with
$\Sigma_{\mathrm{g}}=\Sigma_{\mathrm{HI}}+\Sigma_{\mathrm{H2}}$ and
$\sigma_{\mathrm{g}}=11\;\mbox{km\,s}^{-1}$ (Leroy et al.\ 2008).  What are
the limitations of this approach?  How do H\,\textsc{i} and $\mathrm{H}_{2}$
contribute to star-gas instabilities?  To answer these questions, we consider
H\,\textsc{i} and $\mathrm{H}_{2}$ separately, and choose observationally
motivated values of the 1D velocity dispersion:
$\sigma_{\mathrm{HI}}=11\;\mbox{km\,s}^{-1}$ (Leroy et al.\ 2008), and
$\sigma_{\mathrm{H2}}=6\;\mbox{km\,s}^{-1}$ (Wilson et al.\ 2011).  We then
compute the $(s,q)$ tracks for each case (stars plus H\,\textsc{i} or
$\mathrm{H}_{2}$), and show the results in the right panel of Fig.\ 3.  Note
that H\,\textsc{i} and $\mathrm{H}_{2}$ populate the parameter plane
differently.  In particular, none of the H\,\textsc{i} data falls within the
two-phase region, while $\mathrm{H}_{2}$ populates such a region in 60\% of
the cases.  This means that H\,\textsc{i} and $\mathrm{H}_{2}$ have distinct
stability properties, and a fundamentally different dynamical coupling with
stars.  Treating the ISM as a single component underestimates the role that
$\mathrm{H}_{2}$ plays in star-gas instabilities, and overestimates the
contribution of H\,\textsc{i}.  This is why a multi-component analysis is
needed!

\subsection{Approximating $Q$ in the thin-disc limit}

The only multi-component stability diagnostics that have been available so
far are the local axisymmetric stability criteria of Morozov (1981) and
Rafikov (2001), which are valid for infinitesimally thin discs made of gas
and $N_{\star}$ stellar components.  For $N_{\star}=2$, the case considered
by Morozov (1981), such criteria are equivalent because they are based on
similar approximations.  However, besides being more general, Rafikov's
criterion is simpler and can be written as $Q_{\mathrm{R},N}\geq1$, where
\begin{equation}
\frac{1}{Q_{\mathrm{R},N}}=
\max\{\mathrm{SC}_{\mathrm{R},N}(k)\}\,,
\end{equation}
\begin{equation}
\mathrm{SC}_{\mathrm{R},N}(k)=
\frac{1}{Q_{1}}\,
\frac{2K_{1}}{1+K_{1}^{2}}+
\sum_{i=2}^{N}\frac{1}{Q_{i}}\,
\frac{2\left[1-\mathrm{e}^{-K_{i}^{2}}I_{0}\left(K_{i}^{2}\right)\right]}{K_{i}}\,,
\end{equation}
\begin{equation}
K_{i}=
k\,\frac{\sigma_{i}}{\kappa}\,,
\end{equation}
\begin{equation}
Q_{i}=
\frac{\kappa\sigma_{i}}{\pi G\Sigma_{i}}\,.
\end{equation}

Let us illustrate how to find a simple, accurate, fast and more general $Q$
diagnostic.  In Sect.\ 2.1, we have shown that stars can be accurately
treated as a fluid when evaluating $Q$.  So we can safely replace the kinetic
terms in Eq.\ (12) with their fluid counterparts.  We then face the heart of
the problem: how to estimate the least stable wavenumber, $k_{\mathrm{max}}$,
without the usual maximization procedure.  Consider the two-component case
first, and compare the Romeo-Wiegert approximation [Eqs (9) and (10)] with
the two-fluid stability parameter [Eqs (6) and (7)].  One can easily infer
that $K_{\mathrm{max}}$ fulfils the following conditions:
\begin{enumerate}
\item If $Q_{\star}>Q_{\mathrm{g}}$, then
  $2K_{\mathrm{max}}/(1+K_{\mathrm{max}}^{2})\sim W$ and
  $2K_{\mathrm{max}}s/(1+K_{\mathrm{max}}^{2}s^{2})\sim1$.  As
  $W=2s/(1+s^{2})$, this implies that $K_{\mathrm{max}}\sim1/s$,
  i.e.\ $k_{\mathrm{max}}\sim\kappa/\sigma_{\mathrm{g}}$.
\item If $Q_{\mathrm{g}}>Q_{\star}$, then
  $2K_{\mathrm{max}}/(1+K_{\mathrm{max}}^{2})\sim1$ and
  $2K_{\mathrm{max}}s/(1+K_{\mathrm{max}}^{2}s^{2})\sim W$.  This implies
  that $K_{\mathrm{max}}\sim1$,
  i.e.\ $k_{\mathrm{max}}\sim\kappa/\sigma_{\star}$.
\end{enumerate}
Conditions (i) and (ii) have not been pointed out in previous analyses.  They
simply mean that $1/k_{\mathrm{max}}$ is approximately the typical epicycle
size of the less stable component.  This approximation is not accurate when
$s\approx0.2$ and $q\approx1$, since for such values there is a transition
between three stability regimes and $K_{\mathrm{max}}$ has a jump across one
of the interfaces (see discussion of Fig.\ 3).  Note, however, that
estimating $K_{\mathrm{max}}$ as above produces a very accurate estimate of
$Q_{\mathrm{Rf}}$, namely the Romeo-Wiegert approximation.  This is because
$Q_{\mathrm{Rf}}$ is continuous across the $q=1$ line, and because the error
that affects the estimate of $Q_{\mathrm{Rf}}$ is of second order with
respect to that of $K_{\mathrm{max}}$: $\Delta Q_{\mathrm{Rf}}\propto(\Delta
K_{\mathrm{max}})^{2}$.  The first-order term is obviously zero, since the
first derivative of $\mathrm{SC}_{\mathrm{Rf}}(K)$ vanishes for
$K=K_{\mathrm{max}}$.%
\footnote{This is actually the idea behind all minimization (or maximization)
  problems, and the reason why their solutions are robust.  An instructive
  example is the `optimal' Wiener filter used in signal/image processing
  (see, e.g., Press et al.\ 1992).}
This flow of arguments suggests that we can estimate the $N$-component $Q$
parameter as in the two-fluid case, i.e.\ by approximating
$1/k_{\mathrm{max}}$ with the typical epicycle size of the least stable
component ($k_{\mathrm{max}}\sim\kappa/\sigma_{m}$):
\begin{equation}
\frac{1}{Q_{N}}=
\sum_{i=1}^{N}\frac{W_{i}}{Q_{i}}\,,
\end{equation}
\begin{equation}
W_{i}=
\frac{2\sigma_{m}\sigma_{i}}
     {\sigma_{m}^{2}+\sigma_{i}^{2}}\,,
\end{equation}
where the index $m$ denotes the component with smallest $Q$:
$Q_{m}=\min\{Q_{i}\}$.  This is the component that dominates the local
stability level ($W_{m}=1$).  All other components have less weight; the more
$\sigma_{i}$ differs from $\sigma_{m}$, the smaller the weight factor
$W_{i}$.

What is the accuracy of $Q_{N}$?  In Appendix A, we show how estimation
uncertainties propagate from $W_{i}$ to $Q_{N}$, and derive an upper bound
for the resulting root-mean-square error:
\begin{equation}
\frac{\Delta Q_{N}}{Q_{N}}\la0.03\,\sqrt{N}\,.
\end{equation}
Eq.\ (17) tells us that the accuracy of our approximation deteriorates slowly
as we consider more and more components.  In the case of greatest interest,
stars plus H\,\textsc{i} plus $\mathrm{H}_{2}$, the relative error is on
average less than 6\%, i.e.\ almost as low as in the two-component case (see
Fig.\ 2).

%%%%%%%%%%%%%%%%%%%%%%%%%%%%%%%%%%%%%%%%%%%%%%%%%%%%%%%%%%%%%%%%%%%%%%%%%%%%%%%%%%%%
%%%%%%%%%%%%%%%%%%%%%%%%%%%%%%%%%%%%%%%%%%%%%%%%%%%%%%%%%%%%%%%%%%%%%%%%%%%%%%%%%%%%
\begin{table}
\caption{Accuracy of our approximation in the ten-component cases analysed by
  Rafikov (2001).}
\begin{center}
\begin{tabular}{cccc}
\hline
Case & $Q_{\mathrm{R},N}$ & $Q_{N}$ & $(Q_{N}-Q_{\mathrm{R},N})/Q_{\mathrm{R},N}$ \\
\hline
  1  &        1.10        &   1.12  &                    1.8\%                    \\
  2  &        1.00        &   1.04  &                    3.7\%                    \\
  3  &        1.00        &   1.04  &                    4.5\%                    \\
  4  &        1.00        &   1.01  &                    0.8\%                    \\
 4+1 &        1.00        &   1.01  &                    1.2\%                    \\
\hline
\end{tabular}
\end{center}

\medskip
$Q_{\mathrm{R},N}$ is the $N$-component $Q$ parameter of Rafikov (2001),
$Q_{N}$ is our $N$-component $Q$ parameter, and
$(Q_{N}-Q_{\mathrm{R},N})/Q_{\mathrm{R},N}$ is the relative error of our
approximation.

\end{table}
%%%%%%%%%%%%%%%%%%%%%%%%%%%%%%%%%%%%%%%%%%%%%%%%%%%%%%%%%%%%%%%%%%%%%%%%%%%%%%%%%%%%
%%%%%%%%%%%%%%%%%%%%%%%%%%%%%%%%%%%%%%%%%%%%%%%%%%%%%%%%%%%%%%%%%%%%%%%%%%%%%%%%%%%%

To demonstrate the accuracy of our approximation, we put it to a stringent
test: the 10-component model of the Solar neighbourhood analysed by Rafikov
(2001).  That model includes cold interstellar gas, giants, stars in six
luminosity ranges, white and brown dwarfs.  Rafikov also analysed the effect
of varying the most uncertain model parameters.  He motivated and discussed
five cases, which we summarize in Appendix B.  For each case, we compare
$Q_{N}$ with Rafikov's $Q_{\mathrm{R},N}$ parameter and compute the relative
error $(Q_{N}-Q_{\mathrm{R},N})/Q_{\mathrm{R},N}$.  Table 1 shows that our
approximation is remarkably accurate.  In all cases, the relative error is
less than 5\%, which means twice as low as the upper bound predicted by
Eq.\ (17).

\subsection{Adding the effect of disc thickness}

Our approximation is not yet complete.  It does not include the stabilizing
effect of disc thickness, which is important and should be taken into account
when analysing the stability of galactic discs (Romeo \& Wiegert 2011).  In
this section, we generalize the Romeo-Wiegert approach and provide a simple
recipe for adding such an effect.

From the thin-disc limit, we have learned that the local stability level is
dominated by the component with smallest $Q$ [see Eqs (15) and (16)].  The
contributions of the other components are weakened by the $W_{i}$ factors,
which are different and small if the components are dynamically distinct.  In
this case, we can estimate the effect of thickness reasonably well by
considering each component separately.  Romeo (1994) analysed this case in
detail.  The effect of thickness is to increase the stability parameter of
each component by a factor $T$, which depends on the ratio of vertical to
radial velocity dispersion:
\begin{equation}
T\approx
\left\{\begin{array}{ll}
       {\displaystyle1+0.6\left(\frac{\sigma_{z}}{\sigma_{R}}\right)^{2}}
                       & \mbox{for\ }0\la\sigma_{z}/\sigma_{R}\la0.5\,, \\
                       &                                                \\
       {\displaystyle0.8+0.7\left(\frac{\sigma_{z}}{\sigma_{R}}\right)}
                       & \mbox{for\ }0.5\la\sigma_{z}/\sigma_{R}\la1\,.
       \end{array}
\right.
\end{equation}
Eq.\ (18) can be inferred from fig.\ 3 (top) of Romeo (1994).  The range
$0\la\sigma_{z}/\sigma_{R}\la0.5$ is characteristic of the old stellar disc
in Sc--Sd galaxies (Gerssen \& Shapiro Griffin 2012), while
$0.5\la\sigma_{z}/\sigma_{R}\la1$ is the usual range of velocity anisotropy
(typical of the old stellar disc in Sa--Sbc galaxies, of young stars and the
ISM).  To approximate $Q$ in this more general context, use then Eq.\ (15)
with $Q_{i}$ replaced by $T_{i}Q_{i}$:
\begin{equation}
\frac{1}{\mathcal{Q}_{N}}=
\sum_{i=1}^{N}\frac{W_{i}}{T_{i}Q_{i}}\,,
\end{equation}
where $\mathcal{Q}_{N}$ is our $Q$ stability parameter for multi-component
and realistically thick discs, $Q_{i}=\kappa\sigma_{i}/\pi G\Sigma_{i}$ is
the Toomre parameter of component $i$, $T_{i}$ is given by Eq.\ (18), and
$W_{i}$ is given by Eq.\ (16).  \textbf{NOTE} that the index $m$ now denotes
the component with smallest $TQ$: $T_{m}Q_{m}=\min\{T_{i}Q_{i}\}$.  This is
the component that dominates the local stability level.  The contributions of
the other components are still suppressed by the $W_{i}$ factors.

\subsection{What about the effect of ISM turbulence?}

Turbulence plays a fundamental role in the dynamics and structure of cold
interstellar gas (see, e.g., Elmegreen \& Scalo 2004; McKee \& Ostriker 2007;
Agertz et al.\ 2009).  The most basic aspect of interstellar turbulence is
the presence of supersonic motions.  These are usually taken into account by
identifying $\sigma_{\mathrm{g}}$ with the typical 1D velocity dispersion of
the medium, rather than with its thermal sound speed.  Another important
aspect of interstellar turbulence is the existence of scaling relations
between $\Sigma_{\mathrm{g}}$, $\sigma_{\mathrm{g}}$ and the size of the
region over which such quantities are measured ($\ell$).  Observations show
that $\Sigma_{\mathrm{HI}}\sim\ell^{1/3}$ and
$\sigma_{\mathrm{HI}}\sim\ell^{1/3}$ up to scales of 1--10 kpc, whereas
$\Sigma_{\mathrm{H2}}\sim constant$ and $\sigma_{\mathrm{H2}}\sim\ell^{1/2}$
up to scales of about 100 pc (see, e.g., Elmegreen \& Scalo 2004; McKee \&
Ostriker 2007; Romeo et al.\ 2010).

Motivated by the large observational uncertainties of
$\Sigma_{\mathrm{g}}(\ell)$ and $\sigma_{\mathrm{g}}(\ell)$, and having in
mind near-future applications to high-redshift galaxies, Romeo et al.\ (2010)
considered more general scaling relations,
$\Sigma_{\mathrm{g}}\propto\ell^{a}$ and
$\sigma_{\mathrm{g}}\propto\ell^{b}$, and explored the effect of turbulence
on the gravitational instability of gas discs.  They showed that turbulence
excites a rich variety of stability regimes, several of which have no
classical counterpart.  See in particular the `\emph{stability map of
  turbulence}' (fig.\ 1 of Romeo et al.\ 2010), which illustrates such
stability regimes and populates them with observations, simulations and
models of ISM turbulence.  Hoffmann \& Romeo (2012) extended this
investigation to two-component discs of stars and gas, and analysed the
stability of THINGS spirals.  They showed that ISM turbulence alters the
condition for star-gas decoupling and increases the least stable wavelength,
but hardly modifies the $Q$ parameter at scales larger than about 100 pc.
Since these are the usual scales of interest, we do not include that effect
in our approximation.

%%%%%%%%%%%%%%%%%%%%%%%%%%%%%%%%%%%%%%%%%%%%%%%%%%%%%%%%%%%%%%%%%%%%%%%%%%%%%
\begin{figure*}
\includegraphics[scale=.96]{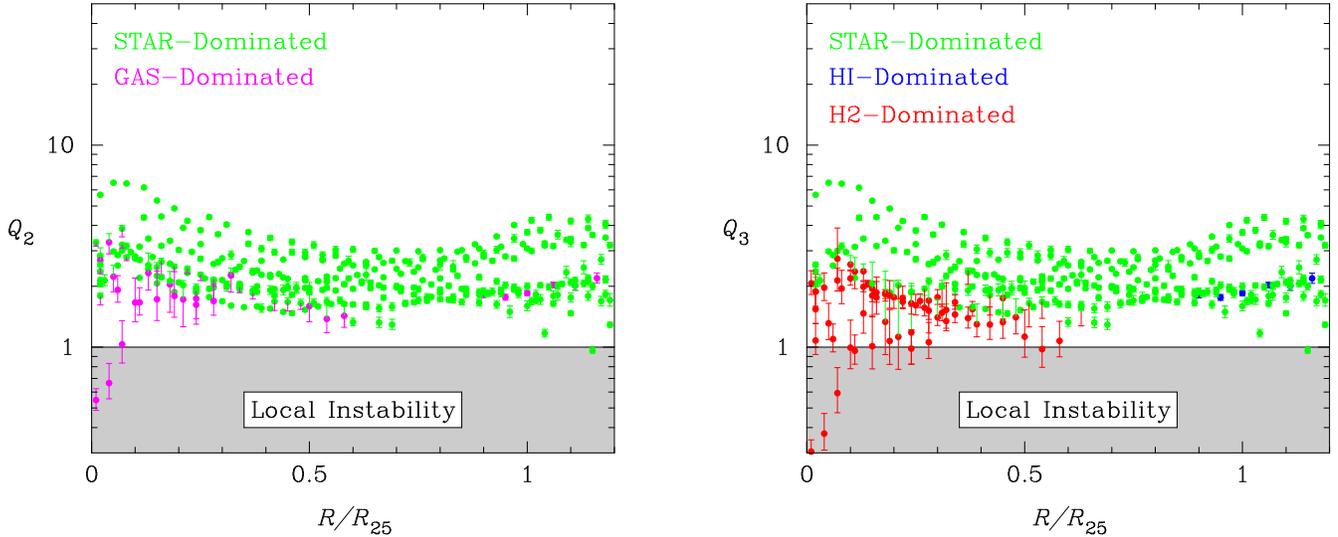}
\caption{The stability level of THINGS spirals.  The diagnostics used in the
  left and right panels are our $\mathcal{Q}_{2}$ and $\mathcal{Q}_{3}$
  parameters [see Eq.\ (19) for $N=2,3$].  The galaxy data are from Leroy et
  al.\ (2008), $R$ is the galactocentric distance, and $R_{25}$ is the
  optical radius.  The data are colour-coded so as to show whether the
  stability level is dominated by stars or gas
  (H\,\textsc{i}/$\mathrm{H}_{2}$), as predicted by Eq.\ (19).  In the right
  panel, the three data points that lie well below the critical stability
  level tell us that the nuclear region of NGC 6946 is subject to strong
  $\mathrm{H}_{2}$-dominated instabilities.  This is consistent with the
  facts that NGC 6946 hosts a nuclear starburst (e.g., Engelbracht et
  al.\ 1996) and a nuclear `bar within bar' (e.g., Fathi et al.\ 2007).}
\end{figure*}
%%%%%%%%%%%%%%%%%%%%%%%%%%%%%%%%%%%%%%%%%%%%%%%%%%%%%%%%%%%%%%%%%%%%%%%%%%%%%

%%%%%%%%%%%%%%%%%%%%%%%%%%%%%%%%%%%%%%%%%%%%%%%%%%%%%%%%%%%%%%%%%%%%%%%%%%%%%
\begin{figure*}
\includegraphics[scale=.96]{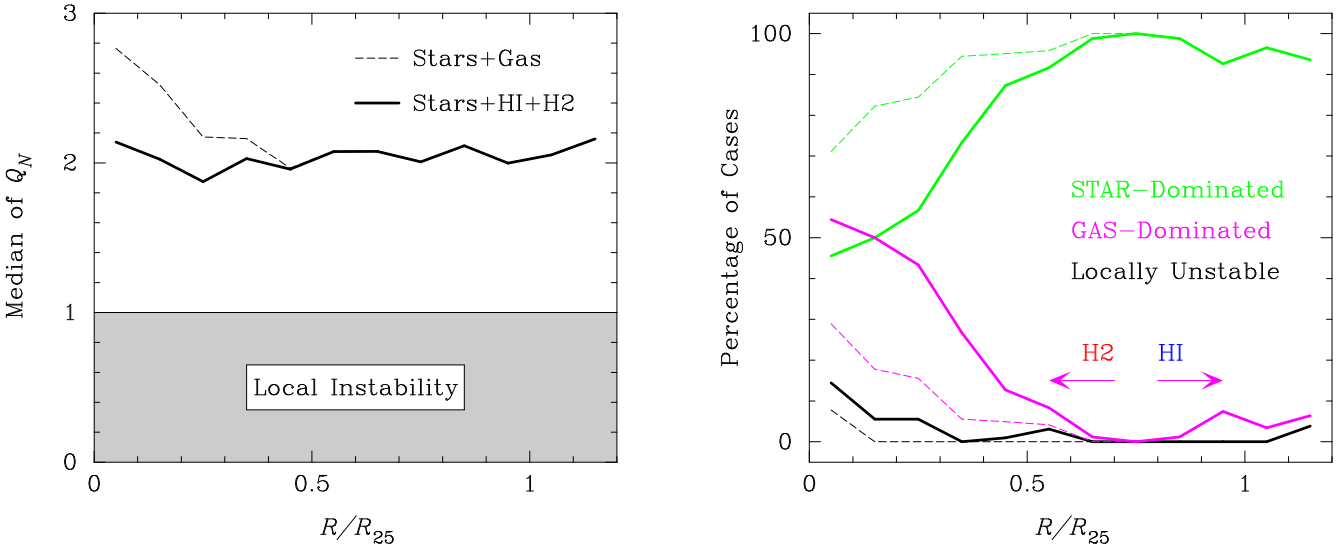}
\caption{Stability characteristics of THINGS spirals.  The radial profiles
  shown in the left and right panels are computed by binning the
  $\mathcal{Q}_{N}$ data of Fig.\ 4 in 12 rings of width $0.1\,R/R_{25}$,
  where $\mathcal{Q}_{N}$ is our $Q$ stability parameter for multi-component
  and realistically thick discs [see Eq.\ (19) for $N=2,3$], $R$ is the
  galactocentric distance, and $R_{25}$ is the optical radius.  The right
  panel shows how frequently the stability level is dominated by stars or gas
  (H\,\textsc{i}/$\mathrm{H}_{2}$), and how frequently $\mathcal{Q}_{N}<1$.
  The thick solid lines are the predictions of a three-component analysis
  (stars plus H\,\textsc{i} plus $\mathrm{H}_{2}$), while the thin dashed
  lines represent the two-component case (stars plus gas).}
\end{figure*}
%%%%%%%%%%%%%%%%%%%%%%%%%%%%%%%%%%%%%%%%%%%%%%%%%%%%%%%%%%%%%%%%%%%%%%%%%%%%%

\subsection{Application to THINGS spirals}

Our approximation is now complete.  Let us then show how to use our
$\mathcal{Q}_{N}$ parameter for analysing the stability of galactic discs,
and illustrate the strength of a multi-component analysis.

In the following, we generalize the two-component approach of Romeo \&
Wiegert (2011).  We consider the same sample of spiral galaxies as in
Sect.\ 2.1 and refer to Leroy et al.\ (2008), hereafter L08, for a detailed
description of the data and their translation into physical quantities.  We
treat stars, H\,\textsc{i} and $\mathrm{H}_{2}$ as three components with the
same surface densities and $\sigma_{\star}$ as in L08, but with distinct
values of $\sigma_{\mathrm{g}}$: $\sigma_{\mathrm{HI}}=11\;\mbox{km\,s}^{-1}$
(L08), and $\sigma_{\mathrm{H2}}=6\;\mbox{km\,s}^{-1}$ (Wilson et al.\ 2011).
For each galaxy, we compute the radial profile of $\mathcal{Q}_{3}$ using
Eq.\ (19).  We adopt $(\sigma_{z}/\sigma_{R})_{\star}=0.6$, as was assumed by
L08, and
$(\sigma_{z}/\sigma_{R})_{\mathrm{HI}}=(\sigma_{z}/\sigma_{R})_{\mathrm{H2}}=1$,
as is natural for collisional components.  For comparison purposes, we also
compute the radial profile of $\mathcal{Q}_{2}$, treating the ISM as a single
component with
$\Sigma_{\mathrm{g}}=\Sigma_{\mathrm{HI}}+\Sigma_{\mathrm{H2}}$ and
$\sigma_{\mathrm{g}}=11\;\mbox{km\,s}^{-1}$ (L08).

Fig.\ 4 shows $\mathcal{Q}_{3}(R)$ and $\mathcal{Q}_{2}(R)$ for the whole
galaxy sample.  Note that the $\mathcal{Q}_{3}$ data are characterized by a
sharp transition at about half the optical radius.  For $R\la0.6\,R_{25}$,
$\mathcal{Q}_{3}$ spans a range of one order of magnitude, and a significant
fraction of the data lie below or near the critical stability level (although
in most of the cases $\mathcal{Q}_{3}>1$).  For $R\ga0.6\,R_{25}$,
$\mathcal{Q}_{3}$ varies within a narrow range of values, and there is a
single data point with $\mathcal{Q}_{3}\leq1$.  Why do the inner and the
outer discs of THINGS spirals have distinct stability properties?  Why does
the transition occur at about half the optical radius?  To answer these
questions, we have colour-coded the $\mathcal{Q}_{3}$ data so as to show
which component dominates the local stability level.  This is an important
piece of information, which can easily be predicted using our
$\mathcal{Q}_{N}$ diagnostic [see Eq.\ (19)].  The fundamental difference
between inner and outer spiral discs is how $\mathrm{H}_{2}$ contributes to
disc (in)stability.  For $R\la0.6\,R_{25}$, $\mathrm{H}_{2}$ dominates in one
third of the cases: it lowers the overall stability level, and increases the
variance of $\mathcal{Q}_{3}$.  At $R\approx0.6\,R_{25}$, the contribution of
$\mathrm{H}_{2}$ becomes negligible.  Thus $\mathrm{H}_{2}$ leaves a
characteristic imprint on the stability of THINGS spirals, even though stars
dominate in most of the cases.  The contribution of H\,\textsc{i} is instead
negligible everywhere, even at the edge of the optical disc, where
H\,\textsc{i} is expected to contribute significantly.  Such a stability
scenario cannot be predicted by a two-component analyis.  Note, in fact, that
the $\mathcal{Q}_{2}$ data underestimate significantly how gas contributes to
disc (in)stability, and fail to reproduce the transition at half the optical
radius (compare the left and right panels of Fig.\ 4).

Fig.\ 5 illustrates how the stability properties of THINGS spirals vary with
galactocentric distance.  To extract such information, we have binned the
$\mathcal{Q}_{3}$ data of Fig.\ 4 in 12 rings of width $0.1\,R/R_{25}$.  For
each ring, we have computed the median of $\mathcal{Q}_{3}$, the percentage
of cases in which $\mathcal{Q}_{3}<1$, and how frequently each component
dominates the local stability level.  For comparison purposes, we have also
binned the $\mathcal{Q}_{2}$ data and computed the corresponding stability
characteristics.  Note that $\mathrm{H}_{2}$ plays a primary role in disc
(in)stability for $R\leq\mbox{0.1--0.2}\;R_{25}$, i.e.\ up to distances of
about one disc scale length ($R_{25}=4.6\pm0.8\;R_{\mathrm{d}}$; L08).
Thereafter stars dominate more often.  Note also that the frequency of
$\mathrm{H}_{2}$-dominated cases decreases markedly with galactocentric
distance, and falls below 10\% at $R=\mbox{0.5--0.6}\;R_{25}$.  This
corresponds to the transition radius identified in Fig.\ 4, and to the
outskirts of the expected $\mathrm{H}_{2}$ domain
($\Sigma_{\mathrm{H2}}\geq\Sigma_{\mathrm{HI}}$ for
$R\leq0.43\pm0.18\;R_{25}$; L08).  The frequency of H\,\textsc{i}-dominated
cases shows a tendency to increase for $R\geq\mbox{0.7--0.8}\;R_{25}$, but it
never rises above 10\%.  The remaining stability characteristics provide more
intriguing information.  The frequency of locally unstable cases is below
10\%, except at distances smaller than about half the disc scale length.  The
median of $\mathcal{Q}_{3}$ lies well above the critical stability level, and
is remarkably constant across the entire optical disc:
$\mathcal{Q}_{3\,\mathrm{med}}(R)\simeq2$ (see the left panel of Fig.\ 5, and
note the linear scale on the $y$-axis).  Note, finally, how fast the
two-component case diverges from the predicted radial profiles as we approach
the galactic centre (compare the dashed and solid lines in the two panels of
Fig.\ 5).  This result illustrates, once again, (i) how important it is to
treat stars, H\,\textsc{i} and $\mathrm{H}_{2}$ as three distinct components
when analysing the stability of galactic discs, and (ii) the strong advantage
of using our $\mathcal{Q}_{N}$ parameter as a stability diagnostic.

\section{DISCUSSION}

Now that we have illustrated the strength of $\mathcal{Q}_{N}$, let us
remember its weaknesses.  Like all $Q$ parameters and stability criteria
discussed so far, $\mathcal{Q}_{N}$ measures the stability of the disc
against local axisymmetric perturbations, so it assumes that $kR\gg1$.  This
is the short-wavelength approximation, which Binney \& Tremaine (2008) define
as `an indispensable tool for understanding the properties of density waves
in differentially rotating disks' (see p.\ 485 of Galactic Dynamics).  Here
the relevant $k$ is the least stable radial wavenumber, which can be
approximated as $k_{\mathrm{max}}\sim\kappa/\sigma_{m}$ (see Sect.\ 2).

%%%%%%%%%%%%%%%%%%%%%%%%%%%%%%%%%%%%%%%%%%%%%%%%%%%%%%%%%%%%%%%%%%%%%%%%%%%%%
\begin{figure}
\includegraphics[scale=.99]{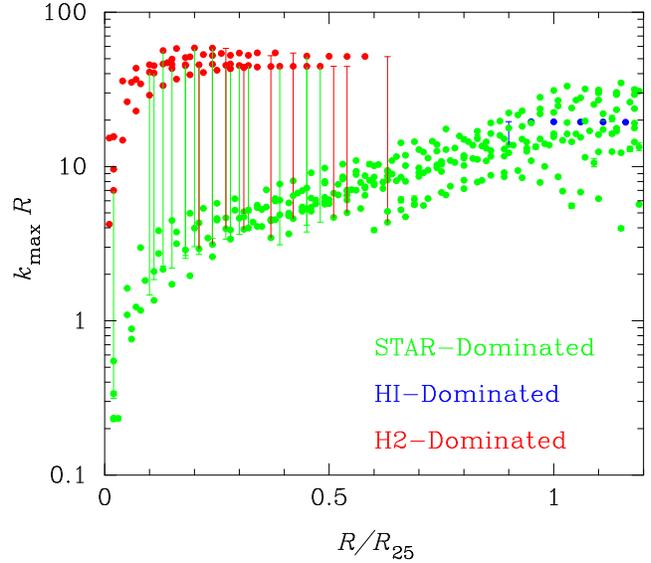}
\caption{Consistency between the results of our stability analysis and the
  short-wavelength approximation, $k_{\mathrm{max}}R\gg1$, where
  $k_{\mathrm{max}}$ is the least stable radial wavenumber, $R$ is the
  galactocentric distance, and $R_{25}$ is the optical radius.  Large error
  bars signal a transition from star- to $\mathrm{H}_{2}$-dominated regimes,
  or vice versa, which causes a jump in the value of $k_{\mathrm{max}}$ (see
  Sect.\ 2).}
\end{figure}
%%%%%%%%%%%%%%%%%%%%%%%%%%%%%%%%%%%%%%%%%%%%%%%%%%%%%%%%%%%%%%%%%%%%%%%%%%%%%

Rather than estimating the magnitude and radial behaviour of
$k_{\mathrm{max}}R$ with qualitative arguments, we plot this quantity as a
function of galactocentric distance for all THINGS spirals analysed in
Sect.\ 2.  Fig.\ 6 illustrates that the results of our stability analysis are
consistent with the short-wavelength approximation.  For instance, the
condition $k_{\mathrm{max}}R>3$ is fulfilled in 93\% of the cases, and is
always true for $R\ga0.25\,R_{25}\approx R_{\mathrm{d}}$, i.e.\ at distances
larger than about one disc scale length (remember that
$R_{25}=4.6\pm0.8\;R_{\mathrm{d}}$; L08).  In contrast, there are only 2\% of
the data with $k_{\mathrm{max}}R\leq1$, all close to the galactic centre:
$R\la0.05\,R_{25}\approx0.25\,R_{\mathrm{d}}$.  Such data correspond to
star-dominated regimes, which are more prone to violate the short-wavelength
approximation since $k_{\mathrm{max}}\sim\kappa/\sigma_{\star}$ and
$\sigma_{\star}>\sigma_{\mathrm{HI}}>\sigma_{\mathrm{H2}}$.  A comparison
with the right panel of Fig.\ 4 shows that the corresponding values of
$\mathcal{Q}_{3}$ are well above unity.  As $\mathcal{Q}_{3}$ is well above
unity and star-dominated in most of the cases, the few data with
$k_{\mathrm{max}}R\leq1$ do not have a significant influence on the stability
of THINGS spirals.  H\,\textsc{i}- and $\mathrm{H}_{2}$-dominated regimes are
fully consistent with the short-wavelength approximation, and so are the
corresponding values of $\mathcal{Q}_{3}$.  In particular, the three data
points that lie well below the critical stability level tell us that the
nuclear region of NGC 6946 is subject to strong $\mathrm{H}_{2}$-dominated
instabilities (see again the right panel of Fig.\ 4).  This is consistent
with the facts that NGC 6946 hosts a nuclear starburst (e.g., Engelbracht et
al.\ 1996) and a nuclear `bar within bar' (e.g., Fathi et al.\ 2007).

While the short-wavelength approximation is satisfied by most spiral
galaxies, the assumption of axisymmetric (or tightly wound) perturbations is
not so general.  Local non-axisymmetric stability criteria are far more
complex than Toomre's criterion: they depend critically on how tightly wound
the perturbations are, and cannot generally be expressed in terms of a single
`effective' $Q$ parameter (e.g., Lau \& Bertin 1978; Morozov \& Khoperskov
1986; Bertin et al.\ 1989b; Jog 1992; Lou \& Fan 1998; Griv \& Gedalin 2012).
However, there is a general consensus that non-axisymmetric perturbations
have a destabilizing effect, i.e.\ a disc with $Q\geq1$ can still be locally
unstable against such perturbations.  Gas dissipation has a similar effect
(Elmegreen 2011).  These may be two of the reasons why the stability level of
THINGS spirals is, on average, well above unity.  The remarkable flatness of
$\mathcal{Q}_{3\,\mathrm{med}}(R)$ across the entire optical disc is far more
intriguing.

The assumption of local perturbations is quite controversial.  While there is
a general consensus that locally stable discs can be globally unstable as
regards spiral structure formation, the dynamics and evolution of spiral
structure depend critically on the radial profile of the $Q$ stability
parameter (e.g., Bertin et al.\ 1989a,\,b; Lowe et al.\ 1994; Romeo 1994;
Korchagin et al.\ 2000, 2005; Khoperskov et al.\ 2007; Sellwood 2011;
Khoperskov et al.\ 2012).  Our results about the stability level of THINGS
spirals have no direct implications for that problem because they concern the
THINGS sample as a whole, not each of the spirals.  Useful constraints on the
nature of spiral structure in galaxies might be found by analysing the radial
profile of $\mathcal{Q}_{N}$ for each THINGS spiral, and by searching for
trends in $\mathcal{Q}_{N}(R)$ along the Hubble sequence.  This is however
well beyond the scope of the present paper.

\section{CONCLUSIONS}

This paper provides a simple analytical recipe for estimating the $Q$
stability parameter in multi-component and realistically thick discs [see
  Eq.\ (19)].  Our $\mathcal{Q}_{N}$ parameter applies for any number of
stellar and gaseous components ($i=1,\ldots,N$), and for the whole range of
velocity anisotropy observed in galactic discs:
$0\la(\sigma_{z}/\sigma_{R})_{i}\la1$.  The accuracy of this approximation
can be rigorously quantified in the thin-disc limit, where it scales as
$N^{-1/2}$.  For $N=3$, the predicted root-mean-square error is well below
10\%.  This is true even for larger values of $N$.  For example, in the
ten-component model(s) of the Solar neighbourhood considered by Rafikov
(2001) our approximation is accurate to within 5\%.  A further strength of
the $\mathcal{Q}_{N}$ diagnostic is that it predicts which component
dominates the local stability level.  This is a useful piece of information,
which should always be given when analysing the stability of galactic discs.

This paper also provides the first three-component analysis of THINGS
spirals.  Our analysis predicts how stars, H\,\textsc{i} and $\mathrm{H}_{2}$
contribute to disc (in)stability, and how the stability properties of such
galaxies vary with galactocentric distance.  We show that $\mathrm{H}_{2}$
plays a primary role up to distances of about one disc scale length.  Stars
dominate thereafter, but the contribution of $\mathrm{H}_{2}$ remains
significant even at distances as large as half the optical radius.  This is
in sharp contrast to the role played by H\,\textsc{i}, which is negligible up
to the edge of the optical disc.  We also show that the stability level of
THINGS spirals is, on average, remarkably flat and well above unity.

\section*{ACKNOWLEDGMENTS}

We are very grateful to Oscar Agertz, Kenji Bekki, Andrew Benson, Andreas
Burkert, Edvige Corbelli, Ed Elson, Frederic Hessman, Volker Hoffmann,
Mordecai-Mark Mac Low, Mathieu Puech, Kyle Westfall, Joachim Wiegert, Tony
Wong and Chao-Chin Yang for useful discussions.  We are also grateful to an
anonymous referee for constructive comments and suggestions, and for
encouraging future work on the topic.  ABR thanks the warm hospitality of
both the Department of Physics at the University of Gothenburg and the
Department of Fundamental Physics at Chalmers.

\appendix

\section{DERIVATION OF EQ.\ (17)}

From standard error analysis, we know that the relative uncertainty of
$Q_{N}$ is approximately equal to that of $1/Q_{N}$:
\begin{equation}
\frac{\Delta Q_{N}}{Q_{N}}\sim\frac{\Delta(1/Q_{N})}{1/Q_{N}}
\end{equation}
(see, e.g., Bevington \& Robinson 2003).  We also know that $\Delta(1/Q_{N})$
arises from estimation uncertainties in $W_{i}$ [see Eq.\ (15)].  If we
assume that all $\Delta W_{i}$ are uncorrelated and of comparable magnitude,
then the error propagation equation reduces to
\begin{equation}
\Delta(1/Q_{N})\sim\Delta W\left(\sum_{i=1}^{N}\frac{1}{Q_{i}^{2}}\right)^{1/2}\,.
\end{equation}
Remembering that $Q_{i}\geq Q_{m}$, we get
\begin{equation}
\Delta(1/Q_{N})\la\frac{\Delta W}{Q_{m}}\,\sqrt{N}\,.
\end{equation}
Noting that $1/Q_{N}>1/Q_{m}$ and using Eq.\ (A1), we then find
\begin{equation}
\frac{\Delta Q_{N}}{Q_{N}}\la\Delta W\,\sqrt{N}\,.
\end{equation}
$\Delta W$ can be evaluated from the two-component case, where the
root-mean-square value of the relative error
$(Q_{2}-Q_{\mathrm{Rk}})/Q_{\mathrm{Rk}}$ is 5\% (see Fig.\ 2).  Setting
$\Delta W\,\sqrt{2}\approx0.05$, we finally obtain $\Delta W\approx0.03$ and
thus
\begin{equation}
\frac{\Delta Q_{N}}{Q_{N}}\la0.03\,\sqrt{N}\,.
\end{equation}

\section{THE TEN-COMPONENT CASES ANALYSED BY RAFIKOV (2001)}

%%%%%%%%%%%%%%%%%%%%%%%%%%%%%%%%%%%%%%%%%%%%%%%%%%%%%%%%%%%%%%%%%%%%%%%%%%%%%%%%%%%%%%%%%%%%%%
%%%%%%%%%%%%%%%%%%%%%%%%%%%%%%%%%%%%%%%%%%%%%%%%%%%%%%%%%%%%%%%%%%%%%%%%%%%%%%%%%%%%%%%%%%%%%%
\begin{table}
\caption{Rafikov's reference model of the Solar neighbourhood: surface
  densities and velocity dispersions of the various components.}
\begin{center}
\begin{tabular}{rcrr}
\hline
$i$ &    Component    & $\Sigma_{i}$ [M$_{\odot}$\,pc$^{-2}$] & $\sigma_{i}$ [km\,s$^{-1}$] \\
\hline
  1 &       ISM       &             13.0   \ \ \ \ \ \ \ \    &           7.0   \ \ \ \ \   \\
  2 &      giants     &              0.4   \ \ \ \ \ \ \ \    &          26.0   \ \ \ \ \   \\
  3 &   $M_{V}<2.5$   &              0.9   \ \ \ \ \ \ \ \    &          17.0   \ \ \ \ \   \\
  4 & $2.5<M_{V}<3.0$ &              0.6   \ \ \ \ \ \ \ \    &          20.0   \ \ \ \ \   \\
  5 & $3.0<M_{V}<4.0$ &              1.1   \ \ \ \ \ \ \ \    &          22.5   \ \ \ \ \   \\
  6 & $4.0<M_{V}<5.0$ &              2.0   \ \ \ \ \ \ \ \    &          26.0   \ \ \ \ \   \\
  7 & $5.0<M_{V}<8.0$ &              6.5   \ \ \ \ \ \ \ \    &          30.5   \ \ \ \ \   \\
  8 &   $M_{V}>8.0$   &             12.3   \ \ \ \ \ \ \ \    &          32.5   \ \ \ \ \   \\
  9 &   white dwarfs  &              4.4   \ \ \ \ \ \ \ \    &          32.5   \ \ \ \ \   \\
 10 &   brown dwarfs  &              6.2   \ \ \ \ \ \ \ \    &          32.5   \ \ \ \ \   \\
\hline
\end{tabular}
\end{center}
\end{table}
%%%%%%%%%%%%%%%%%%%%%%%%%%%%%%%%%%%%%%%%%%%%%%%%%%%%%%%%%%%%%%%%%%%%%%%%%%%%%%%%%%%%%%%%%%%%%%
%%%%%%%%%%%%%%%%%%%%%%%%%%%%%%%%%%%%%%%%%%%%%%%%%%%%%%%%%%%%%%%%%%%%%%%%%%%%%%%%%%%%%%%%%%%%%%

\begin{itemize}
\item \emph{Case 1:} Rafikov's reference model of the Solar neighbourhood.
  The epicyclic frequency is $\kappa=36\;\mbox{km\,s}^{-1}\,\mbox{kpc}^{-1}$.
  The surface densities and velocity dispersions of the various components
  are listed in Table B1.
\item \emph{Case 2:} same as Case 1, but with the velocity dispersion of
  white and brown dwarfs decreased from 32.5 to 20.0 km\,s$^{-1}$.
\item \emph{Case 3:} same as Case 1, but with the total surface density of
  white and brown dwarfs increased from 10.6 to 25.0 M$_{\odot}$\,pc$^{-2}$.
\item \emph{Case 4:} same as Case 1, but with the gas velocity dispersion
  decreased from 7.0 to 5.9 km\,s$^{-1}$.
\item \emph{Case 4+1:} same as Case 1, but with the gas surface density
  increased from 13.0 to 14.8 M$_{\odot}$\,pc$^{-2}$ (this case was only
  mentioned by Rafikov).
\end{itemize}

\bsp

\label{lastpage}

\end{document}